\documentstyle[11pt,paspconf,psfig]{article}
\newcommand{\Msolar}{\mbox{\,$M_{\odot}$\/}}          
\newcommand{\HII}{\mbox{H\,{\footnotesize II}}}       
\newcommand{\oversim}[2]{\protect{\mbox{\lower0.5ex\vbox{%
   \baselineskip=0pt\lineskip=0.2ex
   \ialign{$\mathsurround=0pt #1\hfil##\hfil$\crcr#2\crcr\sim\crcr}}}}} 
\newcommand{\simgreat}{\mbox{$\mathrel{\mathpalette\oversim>}$}} 
\newcommand{\simless} {\mbox{$\mathrel{\mathpalette\oversim<}$}} 
\newcommand{\etal}{\mbox{\hbox{\it et\,al.}}}         
\newcommand{\eg}{\mbox{\hbox{\it e.g.,}}}             
\newcommand{\idest}{\mbox{\hbox{\it i.e.,}}}          
\newcommand{\cf}{\mbox{\hbox{\it cf.}}}               
\newcommand{\kmpers}{\mbox{\,km\,s$^{-1}$}}           
\newcommand{\thetaonec}{\mbox{$\theta ^1$Ori\,C}}     
\def\rhf {R$_{1/2}$}
\def\rh {R_{1/2}}

\def\tcr {t$_{\rm cross}$}
\def\tc {t_{\rm cross}}

\def\tr {t_{\rm relax}}
\def\vdisp {$v_{\rm disp}$}
\def\vdsp {v_{\rm disp}}

\def\msdot {\dot{m}}
\def\ms {m}
\begin{document}
\title{Dynamical Interactions in Dense Stellar Clusters}

\author{Ian Bonnell}
\affil{Institute of Astronomy \\
Madingley Road, Cambridge CB3~0HA, UK \\
bonnell@ast.cam.ac.uk}
\author{Pavel Kroupa}
\affil{Institut f\"ur Theoretische Astrophysik \\
Tiergartenstra\ss{}e 15, D-69121 Heidelberg, Germany \\
pavel@ita.uni-heidelberg.de}
\begin{abstract}
This chapter\footnote{To appear in ASP Conference Series, {\it The
Orion Complex Revisited}, M.J.McCaughrean, A.Burkert (eds.);
International workshop held at Ringberg Castle, June~2--6, 1997}
reviews the dynamical processes that occur in young stellar
clusters. The formation of a stellar cluster is a complex process
whereby hundreds to thousands of stars form near-simultaneously in a
bound configuration. We discuss the dynamical processes involved in
the formation and early evolution of a stellar cluster.

Young clusters are known to contain significant amounts of mass in gas.
The accretion of this gas by individual stars affects the dynamics of
the cluster, and the masses of the stars. Accretion rates depend
crucially on a star's position in the cluster, with those nearer the
centre having higher accretion rates. This results in a spectrum of
stellar masses even if the seeds have equal masses. A by-product of
this process is that the most massive stars are formed in the centre
of the cluster, as is generally observed. Dynamical mass segregation
cannot explain the degree of mass segregation observed in clusters
such as the Trapezium Cluster in Orion, implying that the location of 
the massive stars is an indication of where they formed. This can be 
explained by the competitive accretion model.

Most field stars are found to be members of binary or multiple
systems.  Pre-main sequence stars not in clusters have an even higher
degree of multiplicity, whereas those in dense groupings such as the
Trapezium Cluster have binary frequencies consistent with the field
population. This can be understood if most, if not all, stars form in
binary systems, but that significant numbers of these systems are destroyed
in dense clusters through binary-binary and binary-single
interactions. These models make definite predictions for the
distribution of binary properties. The early evolution of a cluster
sensitively depends on the primordial binary star proportion.

Young stars commonly have circumstellar discs as a remnant of their
formation. Close encounters between stars with circumstellar discs
have drastic effects on the discs and on the stellar orbits.  The
discs are truncated at radii comparable to the encounter periastron,
limiting their lifetimes and affecting their potential for planet
formation.  In small dense clusters, these interactions can leave two
stars bound in a binary system, whereas in larger clusters, the
encounters are too energetic.

Finally, we review the literature on how the removal of residual gas
in the cluster can cause the cluster's dissolution. If the gas
represents a significant fraction of the total mass, its removal on
dynamical timescales can unbind the cluster allowing the stars to
escape and populate the field.
\end{abstract}



%
\section{Introduction}

Recent observations have revealed that most stars do not form singly
and in isolation but rather in groups, from binary systems (\eg{}
Mathieu 1994; Ghez 1995), to small associations (\eg{} Gomez \etal{} 1993), 
to clusters containing hundreds and thousands of stars (\eg{} Lada 
\etal{} 1991; McCaughrean \& Stauffer{} 1994). In star formation regions
such as Orion, it is likely that most stars form in clusters (Lada, 
Strom, \& Myers 1993; Zinnecker, McCaughrean, \& Wilking 1993).
These clusters are typically gas rich, with 50--90\% of their mass 
in gas (\eg{} Lada 1991).

Star formation in such systems is necessarily a complex process,
involving significant dynamical interactions between the stars and 
between the stars and the gas. Unfortunately, that means that
we cannot gain a complete description of the problem from the
commonly used single, isolated star formation scenario
(Shu, Adams, \& Lizano 1987). Instead, we must investigate how these
interactions affect the dynamics of the cluster and their relevance
for explaining the properties of the individual stars. 

The vast amount of recent observational work on young stellar clusters
(\eg{} McCaughrean \& Stauffer 1994; Lada \& Lada 1995; Meyer 1996;
McCaughrean \& O'Dell 1996; Hillenbrand 1997; Hillenbrand \& Hartmann 1998) 
has given an impetus to the theoretical investigations of their formation 
and early evolution. The evolution of young stellar clusters depends on 
the dynamical processes involved, including competitive gas accretion 
by the stars, dynamical mass segregation due to two-body relaxation, 
binary star interactions, star-disc interactions, and finally, the effect 
of gas removal due to its interaction with the constituent high-mass stars.

\section{Preliminaries}

There are currently no satisfactory models capable of explaining the
initial stages of cluster formation. Numerical simulations of
gravitational collapse and fragmentation typically form no more than
10--20 distinct objects (\eg{} Bonnell \etal{} 1992; Burkert \&
Bodenheimer 1993; Boss 1996). This maximum number is due to the
collapse proceeding preferentially in some directions, reducing the
dimensionality of the cloud before fragmentation occurs. This aids in
the fragmentation as sheets and lines are inherently more unstable to
fragmentation than are spheroidals, but limits the degree to which
fragmentation can occur (see the chapter in this volume by Burkert 
\& Bodenheimer).

There are two possible alternatives which might circumvent this problem.  
The first is that the initial cloud contains significant substructure at
the time of collapse (see, \eg{} the chapter by Elmegreen \& Efremov
in this volume) and that the cloud cools rapidly (or is very
gravitationally unstable), so that each ``seed'' contains a Jeans mass
and can thus collapse on its own and more quickly than the overall cloud
(\eg{} Klessen 1997). Such conditions can result from a dynamical
triggering, whereby the cloud is quickly compressed and subsequently
cools in the post-shock layer (\eg{} Whitworth \etal{} 1994; Whitworth \&
Clarke 1997). This is essentially the picture advanced by Fall \& Rees (1985)
to explain the formation of globular clusters. The second possibility is 
that the cluster is formed from an agglomeration of smaller clusters. In 
this case, one would expect some resultant signature of this process in 
the presence of substructure in the cluster. 
Although sub-clustering is evident in some embedded clusters (see, \eg{} 
Lada \& Lada 1995; Lada, Alves, \& Lada 1996; Megeath \& Wilson 1997),
there appears to be none in the Orion Nebula Cluster (ONC) (Bate, Clarke, 
\& McCaughrean 1998; see also the chapter in this volume by Bate \& Clarke), 
but such sub-clustering could have been removed through evolution. 

Subsequent to the initial fragmentation or coagulation event, most
clusters are likely to be significantly out of virial
equilibrium. This follows from considering the pre-cluster entity as
having lost its support (thermal, kinetic, or magnetic) in order to
fragment, and thus the resultant cluster will also lack kinetic energy
and collapse. The ensuing violent relaxation (Lynden-Bell 1967)
revirialises the cluster and produces a large spread in the spectrum
of stellar energies. Violent relaxation is mass-independent, such that
the resultant stellar orbits depend solely on the stars' positions in
the cluster during the event. As it is mass independent, it results in
a near-uniform and mass-independent velocity dispersion, and a
centrally condensed configuration (see Binney \& Tremaine 1987).
Although violent relaxation does erase many of the initial conditions
of the cluster, initial asymmetries in the overall cluster shape can
remain, to a lesser degree, after violent relaxation (Aarseth \&
Binney 1978; Goodwin 1997b). The elongation apparent in the ONC
(Hillenbrand \& Hartmann 1998) may be an indication of such asymmetric 
initial conditions. Cluster contraction is expected even if the 
protostars decouple from the gas in such a way that a cold collapse 
and thus violent relaxation do not occur,
because dynamical friction between the stellar system and the gas
causes deceleration of the stars on a timescale probably much shorter
than the gas expulsion time (Saiyadpour, Deiss, \& Kegel 1997; see also
Gorti \& Bhatt 1996).

Regardless of how the cluster initially formed, for the purpose of
this discussion we consider a cluster as bound (due either solely to
the stars or a combination of stars and gas), and generally spherical
in shape, with no significant substructure present.
The timescale for processes occurring in the cluster are typically
measured in units of the cluster's dynamical or crossing time:
\begin{equation}
\tc\, =\, {2 \times \rh \over \vdsp}~,
\end{equation}
where \rhf{} is the cluster half-mass radius and \vdisp{} is the cluster's
one-dimensional velocity dispersion. This is the shortest timescale
for global cluster evolution, via processes such as violent
relaxation. Sub-regions of the cluster, such as the central core, can
evolve faster and thus independently, on their own crossing time
$t_{\rm core} = 2 \times R_{\rm core} / \vdsp$.  Slower processes,
such as two-body relaxation which drives the cluster towards
equipartition, mass segregation, and core collapse, occur on the
cluster's half-mass relaxation time (Binney \& Tremaine 1987;
see Equation~10 below),
\begin{equation}
\tr \,\approx\, {N \over 8\,\ln\,N}\,\tc~.
\end{equation}
The relaxation time, $\tr$, is a measure of the time it takes for the
kinetic energy, $E_{\rm kin}$, of a star to change by an amount
similar to $E_{\rm kin}$ through many long-distance (\idest{} weak)
encounters with individual stars (\idest{} {\it two-body\/} relaxation),
and thus measures the time it takes for a
stellar cluster to lose memory of its initial dynamical configuration.
Thus larger clusters will take longer to become fully relaxed.  If the
initial phase-space distribution of stars in a cluster consisting of
single stars is independent of stellar mass, then core collapse occurs
typically within 2--3$\,\tr$, and the massive stars with mass
$m_{\rm m}$ sink to the cluster centre on the equipartition timescale
\begin{equation}
t_{\rm eq}\,\approx\,{{\overline m}\over m_{\rm m}}\,\tr~,
\end{equation}
where ${\overline m}$ is the mean stellar mass (Spitzer 1987, p74;
Spurzem \& Takahashi 1995, Equation~A16). Core contraction occurs because,
in a single-star cluster initially in virial equilibrium, the core is
hotter (higher velocity dispersion) than the halo. Redistribution of
kinetic energy (heat conduction) causes the core to {\it cool\/} and the
halo to acquire kinetic energy. A self-gravitating stellar system in
which the velocity dispersion is reduced contracts, leading to a
temperature {\it increase\/}. It has a negative specific heat capacity
(Hachisu \& Sugimoto 1978). The star cluster thus evolves to an
increasingly compact core with an expanding halo. The result is core
collapse (see also Spurzem 1991). Primordial binary stars can, however, 
completely change the evolution (\eg{} Gao \etal{} 1991, and below).

The maximum lifetime of a cluster depends on its size and environment. 
Small clusters will dissolve through interactions with a central binary.  
The timescale for this can be estimated as that required for a binary 
to absorb the total energy of the cluster (Heggie 1974)
\begin{equation}
t_{\rm diss} \,\approx\, {N^2\over 100} \tc~.
\end{equation}
In large clusters, the effects of close interactions with binary
systems are smaller. Instead, stars are removed as a result of energy
transfer from two-body relaxation. The cluster will then dissolve on
its evaporation timescale:
\begin{equation}
t_{\rm evap} \,\approx\, 100\,\tr~, 
\end{equation}
where $\tr$ is the median relaxation time during its lifetime. In
general, a binary-rich cluster with less than or equal to a few thousand
stars will expand due to interactions with binaries that heat the cluster 
(see Section~6.2), increasing the crossing and relaxation times.  Clusters
are not isolated but interact with their surroundings, and can be
disrupted by the galactic tidal field (Terlevich 1987) and through
encounters with molecular clouds (Terlevich 1987; Theuns 1992).
Finally, depending on the shape of the initial mass function, stellar
evolution can significantly compromise the lifetime of a cluster
(\eg{} de\,la\,Fuente Marcos 1996a, 1997). Thus, in general, a realistic
medium-sized cluster will expand until reaching its tidal radius, at
which point it will evaporate by the combined process of evaporation
due to stellar interactions and tidal disruption (Kroupa 1995c).

The ONC contains $\approx 2 \times 10^3$ stars with a mean age of less
than 1\,Myr (Hillenbrand \& Hartmann 1998), has a one-dimensional
velocity dispersion of 2.5\kmpers{} (Jones \& Walker 1988), and a 
half-mass radius of $\approx 0.6$\,pc (Hillenbrand \& Hartmann 1998). 
The relevant timescales are $\tc \approx 4.7 \times 10^5$ years and 
$\tr \approx 30\,\tc$. The cluster age is thus probably 1--3\,\tcr. 
The central core of the ONC, with a radius of $\approx 0.1$\,pc, has 
an age of $\approx 13$ crossing times. If 
${\overline m}\approx 0.5$\Msolar{} and $m_{\rm m}\approx 50$\Msolar{}, 
then $t_{\rm eq}\approx 0.3\,\tc$, so that some
dynamical mass segregation may have occurred (but see below).

\section{Gas Accretion}
Gas accretion has long been known to play an important role in star
formation. Gravitational collapse is a non-homologous process (\eg{}
Larson 1969), forming a small mass core that subsequently accretes the
rest of the surrounding gas. In systems where more than one protostar
is formed, the initial collapse and fragmentation results in the
majority of the mass still in the form of gas (Boss 1986; Bonnell
\etal{} 1991, 1992; Burkert \& Bodenheimer 1993; Bonnell \& Bate 1994). 
Accretion of this gas in binary systems can greatly affect the resultant 
system properties such as the separation, eccentricity, and stellar 
masses (Bate, Bonnell, \& Price 1995; Bate \& Bonnell 1997).

Accretion also plays an important role in the dynamics of a stellar 
cluster. Accretion onto individual stars can significantly affect
their masses, while the combined effect of mass loading and gas drag
decreases the kinetic energy in stars and forces the cluster to
contract. Accretion can profoundly affect the cluster, since most clusters
contain significant fractions of their total mass in gas, and as the
timescale for accretion, probably the gas free-fall time 
\begin{equation}
t_{\rm ff}\,\propto\,\rho_{\rm gas}^{-1/2}~,
\end{equation}
is of the same order as the cluster crossing time 
\begin{equation}
t_{\rm cross}\,\,\propto\,\,\sqrt{{R_{\rm clust}^3} / {(M_{\rm stars} 
   + M_{\rm gas})}} \,\,\propto \,\,(\rho_{\rm gas}+ \rho_{\rm stars})^{-1/2}~.
\end{equation}

In a previous symposium on the Orion Nebula, Zinnecker (1982) showed
how competitive accretion, where the individual accretion rates depend 
solely on the square of the star's mass ($\msdot \propto \ms^2$; Equation~8 
below), can lead to a spectrum of stellar masses from a small variation 
in the initial distribution. Although this study neglected the gas 
dynamics, it did point out the possibilities of explaining the large 
range in stellar masses through a simple physical process: accretion.

In a recent study (Bonnell \etal{} 1997), the gas dynamics was
included in the competitive accretion model. This was done by
modelling the stars and gas in the cluster with a hybrid NBODY-SPH
code (Bate \etal{} 1995). This allows the gas dynamics, the stellar 
dynamics, the gas accretion by individual stars, and the changing 
cluster potential to be modelled self-consistently during the evolution.

Accretion affects the stellar dynamics in two ways. Firstly, the
accreted mass will have, in general, zero net-momentum compared to the
stellar orbits. It will thus decrease the stars' kinetic energies while
conserving momentum. Secondly, the stars will excite wakes in the gas,
and thus lose kinetic energy via the associated gas drag. Both processes
will combine to make the cluster more bound, with the stars that
accrete the most (they will also excite larger wakes) sinking more
rapidly to the centre.

The primary result of Bonnell \etal{} (1997) is that the accretion rates
depend crucially on the stars' positions in the cluster. Stars nearer
the centre of the cluster are able to accrete significantly more than
are those that are in the outer regions of the cluster. This is
illustrated in Figure~1, showing the radius in the cluster and the
relative mass accretion rate for a cluster of ten stars. The stellar
masses are initially uniform and the emerging mass spectrum is solely
due to the competitive accretion process. Three stars are highlighted
in order to illustrate how the accretion depends on the star's location.

\begin{figure}[t]
\centerline{\psfig{figure=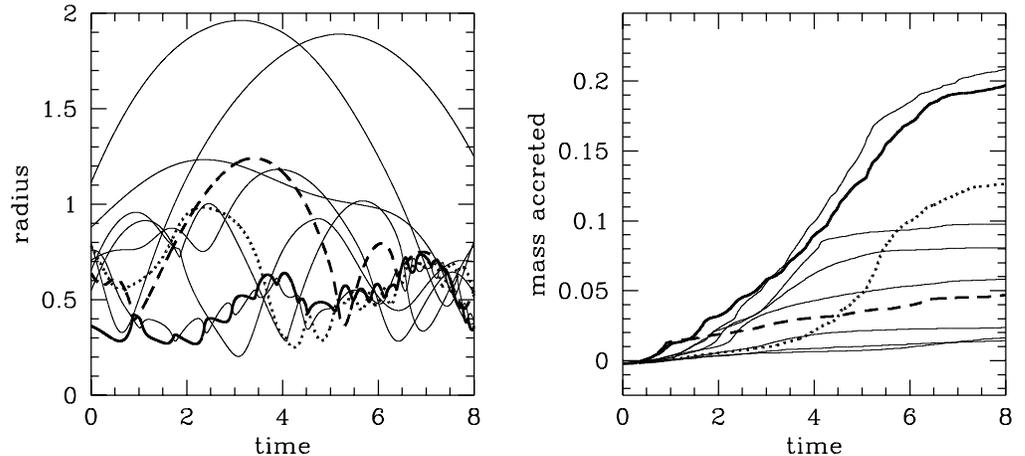,width=\textwidth}}
\caption{\label{cluacc} The evolution of a cluster of ten stars
undergoing gas accretion.  The distance of each star from the
centre of the cluster (left panel) and the gas mass fraction accreted by
each star (right panel) are given as functions of time, in units where
\tcr\,=\,2.8. The gas initially comprises 10\% of the total mass of the
system. Three individual stars are highlighted (see text).}
\end{figure}

The stars closest to the centre of the cluster (heavy solid line in
Figure~1) accrete more than do the other stars because of the increased
gas density near the centre. The density is greater because the gas
preferentially falls into the deepest part of the cluster's
gravitational potential. Furthermore, gas is efficiently funnelled
down to the cluster centre, thereby replenishing the gas as it is
accreted.  Stars that spend most of their time in the outer regions of
the cluster do not have this benefit so they do not accrete nearly as
much gas. This is shown by the heavy dashed line, where the
star is ejected early on from the central regions of the cluster and
does not accrete much gas thereafter. The heavy dotted curve
shows how a star that initially is in the outer regions of the
cluster does not accrete much gas until it comes near the centre of
the potential. The stars which stay on the outside of the cluster (top
curves of left panel of Figure~1) never accrete appreciable amounts of
gas (lowest curves on right panel).

\begin{figure}[t]
\centerline{\psfig{figure=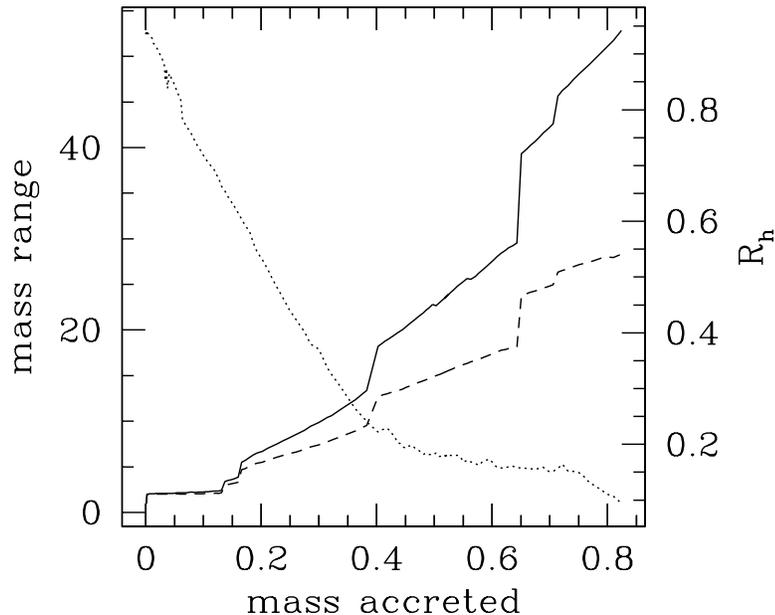,width=4in}}
\caption{\label{massrange} The mass range (defined as the maximum over
the minimum mass) for a cluster of 100 stars as a function of the
total accreted mass in units of the total initial stellar mass (solid
line). Also shown is the ratio of maximum to median mass (dashed line)
and the half-mass radius (dotted line), in units of the initial
half-mass radius. The gas initially comprises 90\% of the
system mass.}
\end{figure}

Accretion in a cluster can thus produce a large range in stellar
masses from an initially uniform distribution. Figure~2 shows the
ratio of maximum to minimum mass and the ratio of the maximum to
average mass, as a function of the amount of gas accreted (in units of
the total initial stellar mass) for a cluster of 100 stars accreting
from gas that initially comprises 90\% of the total mass
(Bonnell \& Bate, in preparation). The range in stellar masses becomes
very large ($m_{\rm max}/ m_{\rm min} \approx 50$), even though the
average stellar mass has only increased by a factor two, and the
cluster has only accreted $\approx 10$\% of the gas. The rapid jumps
in the mass range are due to collisions between individual stars which
are assumed to occur if they pass within 0.1\,AU of each other.  The
cluster is initially contained within 0.1\,pc. Thus, the simple
physical process of competitive accretion in a stellar cluster can
produce a spectrum of stellar masses and thus possibly account for the
observed initial mass function (Scalo 1986; Kroupa, Tout, \& Gilmore 1993).

The competitive accretion process results in the most massive stars 
forming near the centre of the cluster. The stars that
happen to be near the centre accrete more than the rest and thus
it is these that end up as the most massive. This naturally explains the
predominance of massive stars being found near the centre of 
rich clusters. Other explanations for their being located there pose
serious problems (see below). 

The resultant spectrum of stellar masses will thus predominantly relate 
to the initial distribution of stars and gas in a cluster, and then to
the competitive accretion process between the massive stars located
near the cluster centre.  Although it is impossible to calculate the
exact form of the resultant stellar mass spectrum without following a
sufficient number of evolutions, we can make certain qualitative
predictions based on the assumed distributions and the accretion
physics assuming near Bondi-Hoyle accretion (\eg{} Ruffert 1996)
\begin{equation}
\msdot \,=\, 4 \pi \rho \,{(G \ms)^2 \over (v^2 + c_s^2)^{3/2}}~.
\end{equation}
First of all, assuming that the overall distribution is that of a
globally static (non-collapsing) isothermal sphere 
($\rho$$\,\propto$$\,r^{-2}$), that the stars and gas trace the 
same regions in the same fashion, and that the accretion is completely 
mass independent, then the resultant mass spectrum would be 
$dN/dm \propto m^{-3/2}$.
Alternatively, if the accretion depends solely on the star's mass
(Equation~8 with uniform density as in the cluster core), then the
resultant spectrum is the same as found by Zinnecker (1982),
\idest{} $dN/dm \propto m^{-2}$. Generally, accretion will depend 
on both the stellar mass and the gas density so that the initial 
mass function power-law index will be a combination of the above.

\section{Dynamical Mass Segregation}
An alternative explanation for the location of the massive stars
preferentially near the centre of a cluster is that they formed
elsewhere in the cluster and subsequently sank to the centre through
two-body relaxation. It is well known that this process results in
mass segregation, as the encounters redistribute kinetic energy from
the more massive to the less massive stars, driving the system towards
energy equipartition. The question is, can it occur quickly enough to
explain the central location of massive stars such as the Trapezium
OB stars in the ONC within its young age (Equation~3)?

\begin{figure}[t]
\centerline{\psfig{figure=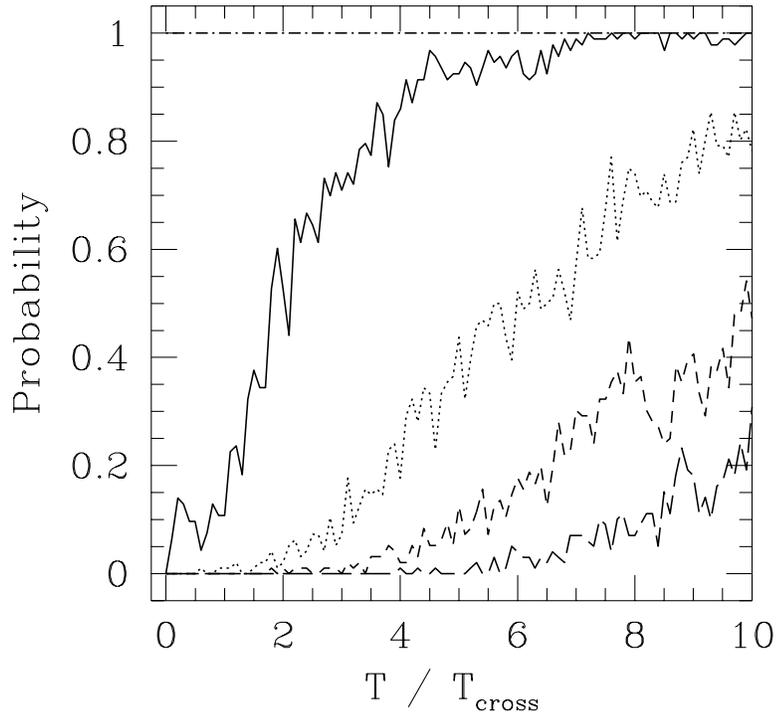,width=4.0in}}
\caption{The probability of observing a Trapezium-like system at the 
centre of a 1500 member cluster as a function of the time in units of the 
crossing time. The ONC is probably on the order of three crossing times 
old or younger. The five lines represent the cases when the 30 most
massive stars are placed (from top to bottom): at Lagrangian radii
containing 2\% (\idest{} the 30 massive stars are the centremost
stars from the beginning), 10\%, 20\%, 30\%, and 50\% of the stars in 
the cluster.}
\label{trapprob}
\end{figure}

Dynamical mass segregation in young stellar clusters has recently been
investigated (Bonnell \& Davies 1998). Although some degree of mass
segregation occurs relatively quickly in most clusters (Equation~3),
it requires of order a relaxation time to be fully mass segregated. As
most young clusters are significantly younger than a relaxation time
(\eg{} the ONC is $\simless\,(1/10)\,\tr$), we can constrain the
initial locations of the massive stars.

Using recent observational studies (McCaughrean \& Stauffer 1994;
Hillenbrand 1997; Hillenbrand \& Hartmann 1998) of the Orion Nebula
cluster, Bonnell \& Davies (1998) showed that the massive stars that
constitute the Trapezium, the central dense core, need to have formed
near where they are presently located, and cannot have evolved there
due to dynamical mass segregation. The simulations of binary-rich
dense clusters lead to the same conclusion (Kroupa, Petr, \&
McCaughrean, in preparation). If the massive stars originated elsewhere 
in the cluster, then the mean stellar mass in the core would be significantly
lower than that observed ($\approx 5$\Msolar{} for the inner 25--40
stars). Furthermore, the probability of forming a dense grouping
containing a significant fraction of the most massive stars (the
Trapezium) is very small unless these massive stars originated nearby
(Figure~3).

We are left with the conclusion that the most massive stars need to have 
formed in the centre of the cluster. Furthermore, it is unlikely that 
they formed from a direct collapse and fragmentation, as the physical 
conditions in the centre of a cluster would lead to the preferential 
formation of low-mass stars there (Zinnecker \etal{} 1993; Bonnell \& 
Davies 1998; Bonnell, Bate, \& Zinnecker 1998). Basically, the high stellar 
density implies a high pre-fragmentation gas density, which in turn implies 
a low fragment mass. Thus the most probable formation mechanism for 
high-mass stars is the competitive accretion process described above, 
possibly combined with accretion induced collisions (see below).

\section{Collisions and the Formation of Massive Stars}
Although accretion in young clusters results in a large range of
stellar masses, it may not be able to explain the formation of massive
($m \simgreat 10\Msolar$) stars (Yorke 1993).  This is because the large
radiation pressure from these stars, combined with the opacity of
interstellar matter, can result in the halting of any further accretion 
(see chapter in this volume by Yorke \& Zinnecker). An alternative formation 
mechanism arises due to the dynamics of an accreting cluster.

Accretion onto a rich young cluster will force it to contract significantly, 
on the timescale of \tcr{} (see Figure~2). The added mass increases the 
binding energy of the cluster, while accretion of basically zero momentum 
matter will remove kinetic energy.  If the core is sufficiently small 
that its crossing time is relatively short compared to the accretion 
timescale, then as shown by Bonnell \etal{} (1998), it will contract 
with added mass as 
\begin{equation}
R_{\rm core} \,\propto\, M_{\rm core}^{-3}~.
\end{equation}
This increases the stellar density dramatically, to the point where
collisions become important.  Collisions between intermediate mass
stars ($2\Msolar\simless m \simless 10 \Msolar$), whose mass has been
accumulated through accretion in the cluster core, can then result in
the formation of massive ($m \simgreat 50 \Msolar$) stars (Bonnell
\etal{} 1998). In this model, the massive stars that are found in the
centre of rich young clusters are significantly younger than the mean
stellar age. If this is how the massive stars in the Trapezium formed,
then we would predict that \thetaonec{} is significantly younger than
$10^6$ years, possibly as young as $\simless 10^5$ years. Such a
young age has implications for structures that are ionised by
\thetaonec, \eg{} the \HII{} region and the proplyds (see the chapters
by Bally, O'Dell, and Johnstone \& Bertoldi).


\section{Binary Stars}
Binary systems in young stellar clusters can play an important role
due to their relatively large cross section for interactions.  In this
way they can both affect the dynamics of clusters and provide insights
into the formation environments of most stars.  Surveys of
main-sequence stars have ascertained that a significant fraction
(typically $\simgreat\,50$\%) are in binary systems (\eg{} Duquennoy \&
Mayor 1991). Observations of sparse groups of pre-main sequence stars
in Taurus-Auriga imply a significant overabundance of binary systems
relative to the main sequence population (\eg{} Ghez 1995). An important
question that any theory of star formation needs to answer is whether
star formation naturally leads to a large binary proportion under all
conditions (Clarke 1995), and if so, how this relates to the
proportion of binaries found on the main sequence.  This has important
implications for the origin of the Galactic field population and the
fraction of which may have formed in the extreme star-forming
environments found in the dense cores of very young clusters.

The best available observational constraints on the abundance of
pre-main sequence binary systems in the core of a very dense and very
young cluster are those of Petr \etal{} (1998; see also the chapter 
by Petr \etal{} in this volume). They applied near-infrared speckle 
holography to stellar systems in the core of the Trapezium Cluster, 
which in turn can be considered the core region of the more massive 
and extended ONC (Hillenbrand \& Hartmann 1998). They found that the 
binary proportion for low-mass Trapezium Cluster stars is indistinguishable 
from the binary proportion of low-mass Galactic field main sequence stars 
over the range of separations 63--225\,AU\@. The binary proportion of
low-mass stars within a radius of 0.25\,pc of the centre of the
Trapezium Cluster was found by Prosser \etal{} (1994) also to be
consistent with the Galactic field value, which is confirmed by an
extended study by Petr \etal{} (this volume). These observations thus
indicate that the binary proportion may be independent of the radial
distance within 0.25\,pc of the centre of the Trapezium Cluster.

It is of considerable interest to investigate where there is a difference 
between the binary proportions of the Trapezium Cluster and Taurus-Auriga.  
Such a difference may be a result of star-formation in low-temperature 
clouds (isolated star formation, \eg{} Taurus-Auriga), and higher-temperature 
clouds (embedded clusters, \eg{} the Trapezium Cluster), as has been 
suggested by Durisen \& Sterzik (1994). A temperature sensitivity may 
be responsible for the difference between Taurus-Auriga and the Galactic 
field, if most stars form in warm molecular clouds and thus in embedded 
clusters. However, the dependence on cloud temperature is by no means an 
established fact at present, owing to the great complexity of the physics 
of star formation.

Another mechanism that decreases a primordial binary proportion is the
disruption of binary systems through binary-binary and binary-single
star encounters (Kroupa 1995a, 1995b; de\,la\,Fuente Marcos 1996b).  
The central 0.1\,pc diameter core of the Trapezium Cluster has roughly
$4.7\times 10^4$ stars pc$^{-3}$ (McCaughrean \& Stauffer 1994),
making this region an environment with extraordinary high stellar
density. The inter-stellar distances amount to about 6000\,AU\@. For
a binary system with a total mass of 1\Msolar, this corresponds to
an orbital period, $P$ (in days), of $\log P\,=\,8.2$. Binary systems
with orbital periods similar to or larger than this value will be
disrupted due to crowding. Furthermore, binary systems with binding
energies comparable to the kinetic energy of a perturber will be
disrupted. The one-dimensional velocity dispersion within the central
radius of 0.41\,pc is about 2.5\kmpers{} (Jones \& Walker 1988). A binary
system with a system mass of 1\Msolar{} and with a circular orbital
velocity equal to the velocity dispersion, has an orbital period
$\log P_{\rm th}\,=\,5.8$, where $P_{\rm th}$ is a ``thermal'' period
corresponding to the ``heat'' of the cluster's internal motion. 
For a binary system with potential energy $E_{\rm pot}$ and orbital 
kinetic energy $E_{\rm orb}$, the binding energy 
$E_{\rm bin}=E_{\rm pot}/2=E_{\rm pot}+E_{\rm orb}$, so that
$E_{\rm orb}=-E_{\rm bin}$.  Thus, most binary systems with
$\log P\,>\,\log P_{\rm th}$ will be disrupted, because an incoming 
perturber can transfer enough kinetic energy to the system to make it unbound.

Assuming the Trapezium Cluster stars have an age of about $\simless 1$\,Myr 
(Prosser \etal{} 1994; Hillenbrand 1997), a stellar system will
have had time to cross a distance of 0.2\,pc (the central core)
$\approx 13$ times. Thus there will have been many interactions
between stellar systems in the cluster core despite its relative youth.
Analytic treatment of the outcomes of binary-binary and
binary-single star interactions, and of the irregular perturbations
acting on the orbit of a binary system in the tidal field of a dense
cluster, is non-trivial (Heggie 1975; Heggie \& Rasio 1996). In order
to estimate the outcomes and their cross-sections, the analytical
treatment has to be augmented by a very large number of
numerical scattering experiments.

Such investigations show that binary systems with separations greater
than that corresponding to the above thermal limit (soft binaries) are
disrupted, while those with smaller separations (hard binaries) tend to
become more bound. To compensate for the increase in binary-star binding
energy, the kinetic energy of the cluster field population
increases. Sufficiently bound binaries are thus an energy source in a
stellar cluster and cause it to expand. Disruption in binary-single
star encounters of binary systems with a period somewhat larger than
the thermal limit cools the cluster, because a part of the kinetic
energy of a perturber is used to ionise the binary system.  This
causes cluster contraction. Disruption of a binary in a binary-binary
encounter, on the other hand, can lead to the formation of a more
bound binary and two ejected single stars. If these do not escape from
the cluster then such four-body collisions effectively heat the
cluster.  Thus there is competition between cooling and heating due to
binary star activity, which operates within the first few tens of
crossing times, until most soft binaries are disrupted.  The
probability of stars merging is enhanced significantly in
binary-binary collisions (Bacon, Sigurdsson, \& Davies 1996), and
stellar evolution complicates the stellar interactions in close
encounters (\eg{} Portegies Zwart 1996; Portegies Zwart \etal{} 1997).

Exchange reactions near the cluster core lead, on average, to the
least-massive star being ejected from the temporary few-body
systems. Binary systems will suffer a recoil from interacting with
single stars and may even be ejected if the interaction does not force
them to merge (Davies, Benz, \& Hills 1994). Tightly bound binaries
are usually heavier than single stars, and will accumulate in the core
through dynamical mass segregation. The energy production through
hardening binaries leads to an expanding cluster core, and the
innermost cluster region is depleted of low-mass stars. Reviews of
these processes can be found in Hut \etal{} (1992), Davies (1995),
Meylan \& Heggie (1997), and also Giannone \& Molteni (1985).

In general, the binary fraction can be significantly reduced by
stellar interactions, with the resultant binary fraction being
dependent on the hard/soft binary separation. This limiting separation
depends on the velocity dispersion in the cluster and hence its
density. Thus, stellar interactions will result in a lower binary fraction
in denser clusters (see Figure~4).

\begin{figure}[t]
\centerline{\psfig{figure=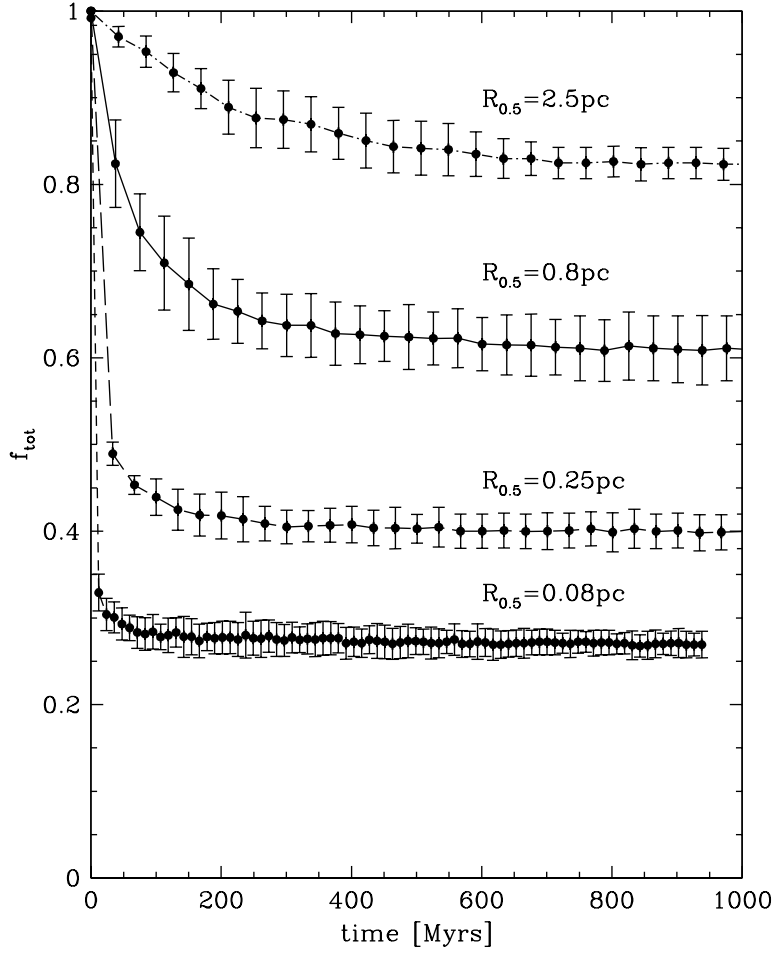,width=4.0in}}
\caption{\label{fbinev} The binary fraction in clusters of different 
half-mass radii versus time. Each cluster initially contains 200 
binary systems.}
\end{figure}

The timescales for such processes are too long (many initial
relaxation times) to noticeably affect the dynamics of the young ONC
on a global scale. For example, it takes about 400\,Myr
for the mean stellar mass inside the central sphere with a radius of
2\,pc to increase by 30\%, for a cluster of initially 200 binary
stars with an initial half-mass radius of 0.08\,pc (Kroupa 1995c). 
However, it is the core of a stellar cluster where most ``action'' 
occurs, and it reacts on a significantly shorter
timescale. The timescales for changes of its structure are,
however, poorly known, mostly because there have only been a few
investigations of clusters that have initially a large proportion of
binary stars. A convenient estimate of such a timescale is the
half-mass relaxation time, which for a stellar system in virial
equilibrium that consists of $N$ stars with mean mass $\overline{m}$,
half-mass radius $R_{1/2}$, and total mass $M_{\rm cl}$, is
(from Equation~2):
\begin{equation}
t_{\rm relax}\,=\,{2.1\times10^7\,{\rm yr}\over \ln(0.4\,N)}
   \left({M_{\rm cl} \over 100\Msolar}\right)^{1/2} 
   \left({1\Msolar\over \overline{m}}\right) 
   \left({R_{1/2} \over 1\,{\rm pc}}\right)^{3/2}~.
\end{equation}

A stellar system with $R_{1/2}$\,=\,0.1--0.3\,pc, $N=500$, and 
$M_{\rm cl}=300$\Msolar, has a relaxation time 
$t_{\rm relax}$\,=\,0.7--3.7\,Myr. These values are characteristic for 
the Trapezium Cluster, the core of the ONC\@. Also, for the Trapezium 
Cluster, $\sigma \approx 2.5$\kmpers, and so 
$t_{\rm cross}$\,=\,4--12$\times 10^4$\,yr. 
A typical star may thus have crossed the central region many times, as
stated above. This implies that collisions between stellar systems
will have been frequent in the central few tenths of a parsec of the
Trapezium Cluster.

The considerations presented above suggest that the Trapezium Cluster
may be old enough to perhaps allow an initial Taurus-Auriga binary 
population to have dynamically evolved to the observed reduced value. 
Whether this is true can be investigated by numerical simulations.  
This is the subject of the following sub-sections, further details of
which will be available in Kroupa, Petr, \& McCaughrean, in preparation. 

\subsection{A model Trapezium Cluster} \label{ic}
The ONC is the presently best studied very young and very dense cluster. 
Furthermore, the central core, the Trapezium Cluster, is sufficiently 
small that it may be a dynamically evolved structure.  It is thus 
sensible to take the Trapezium Cluster as a starting point for the 
study of the dynamical processes involving young binary stars.

\begin{figure}[t]
\centerline{\psfig{figure=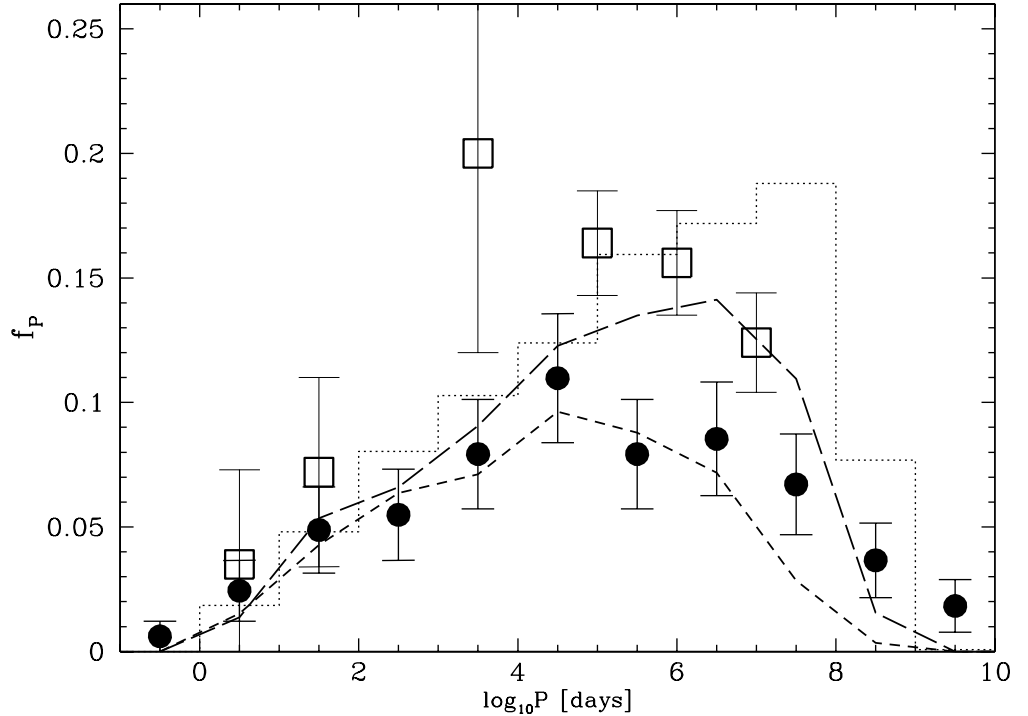,angle=270,width=\textwidth}}
\caption{Distribution of orbits, $f_P$, for main sequence multiple
systems (solid dots, Duquennoy \& Mayor 1991) and pre-main sequence
systems in Taurus-Auriga (open squares: $\log P$\,$>$\,4, K\"ohler \&
Leinert 1998; $\log P$\,=\,3.5, Richichi \etal{} 1994; $\log P$\,$<$\,2,
Mathieu 1994). Main sequence G-, K- and M-dwarf binaries have
essentially the same period distribution (\cf{} Figure~1 of Kroupa 
1995a). The dotted histogram is the initial period distribution from Kroupa
(1995b, Figure~7). Crowding in the model Trapezium Cluster changes this
distribution to the long-dashed one, which is Model~A\@. A Gaussian
log-period birth distribution that fits the solid dots, changes
through crowding in the model Trapezium Cluster to the distribution
shown as the short-dashed line. This is Model~B\@.}
\label{fig:init_P}
\end{figure}

The following parameters are chosen to specify a model Trapezium
Cluster at birth, \idest{} at the time when stellar dynamics starts
dominating over gas dynamics: 
\begin{itemize}
\item $N = 1600$ stars
\item An initial mass function deduced from star-count data in the Galactic 
      field by Kroupa, Tout, \& Gilmore (1993), based on Scalo's (1986)
      determination for $m>1\,M_\odot$).  
\item Lower and upper stellar mass limits, $m_{\rm l}$ and $m_{\rm u}$,
      of 0.08 and 30\Msolar{} respectively 
\item A Plummer density distribution in virial equilibrium with
      $R_{1/2}=0.1$\,pc 
\item Initial position vectors and velocity dispersion independent of 
      stellar mass
\item An isotropic velocity distribution for the binary centres-of-mass 
\end{itemize}
The resulting cluster mass is $M_{\rm cl}=700$\Msolar. The initial
number of stars is guided by the detection in the near-infrared of 500
probable Trapezium Cluster members within a radius of about 0.65\,pc
(Zinnecker \etal{} 1993). Probably not all Trapezium Cluster stars 
will have been detected, and most binary systems would have not been
resolved in that study. Also, allowance must be made for the dynamical
evolution during the first Myr. The initial half-mass radius is
guided by the modelling of McCaughrean \& Stauffer (1994).  These
parameters are a compromise, in that they give a central density that
is larger by an order of magnitude than the observed value, and a
velocity dispersion that is somewhat smaller than the observed value.
The relaxation and crossing times are $t_{\rm relax}=0.62$\,Myr and
$t_{\rm cross}=0.1$\,Myr, respectively. The central number density will
decrease and the half-mass radius will increase within 1\,Myr owing
to the heating through binary stars. The assumption that the velocity
and position vectors are not correlated with the stellar mass is 
contradicted by the observations, which show that the massive stars
are concentrated in the cluster core. However, the present assumption
allows investigation of whether dynamical mass-segregation in young
binary-rich clusters can lead to the observed mass-segregation on
timescales of 1\,Myr or less (see above).
For the primordial binary star population two models are investigated: 
\begin{itemize}
\item Model~A: All stars are in $N_{\rm bin}=800$ binary systems;  
      the initial period distribution is taken from Kroupa (1995b), with
      $\log P_{\rm min}\approx 0$ (0.02\,AU for a 1\Msolar{}
      system) and $\log P_{\rm max}=8.43$ (8200\,AU for a 
      $1\,M_\odot$ system)
\item Model B: 1200 stars are in $N_{\rm bin}=600$ binary systems, and 
      $N_{\rm sing}=400$ stars are single; a Duquennoy \& Mayor (1991) 
      initial log-period distribution, with $\log P_{\rm min}=0$ and 
      $\log P_{\rm max}=11$ ($4.2\times10^5$\,AU for a 1\Msolar{} system)
\end{itemize}
In Model~A, the binary fraction, $f_{\rm tot}$, is 1, consistent with 
the pre-main sequence binary proportion in Taurus-Auriga. In Model~B, 
$f_{\rm tot}=0.6$, consistent with the binary proportion for Galactic
field main sequence stars. Here, $f=N_{\rm bin}/(N_{\rm bin}+N_{\rm sing})$, 
and the subscript ``tot'' implies that all periods are counted.
The initial model period distributions are compared with the
observational data in Figure~\ref{fig:init_P}.
In addition, the following assumptions are made: the initial
mass-ratio distribution is random and the initial eccentricity
distribution is thermally relaxed (Kroupa 1995a, 1995b). All results
quoted here are averages of three equivalent simulations per model.


Computer simulations of stellar clusters are problematic because
the forces have to be computed for each pair of stars, so that the
computational time scales as $N^2$. Binary stars, some of which have
to be integrated with very small time-steps, cause additional serious
delays. An $N$--body programme must allow the simulation of dynamical
processes that have timescales ranging from days to many $10^8$\,yr.
For this purpose, the best available code for the realistic simulation
of stellar clusters, {\sc Nbody5}, was used (Aarseth 1985, 1994). It
incorporates many special algorithms to ensure computational speed and
efficiency without compromising accuracy. Short descriptions of the
code may be found in Hut \etal{} (1992) and Meylan \& Heggie (1997).


\begin{figure}[t]
\centerline{\psfig{figure=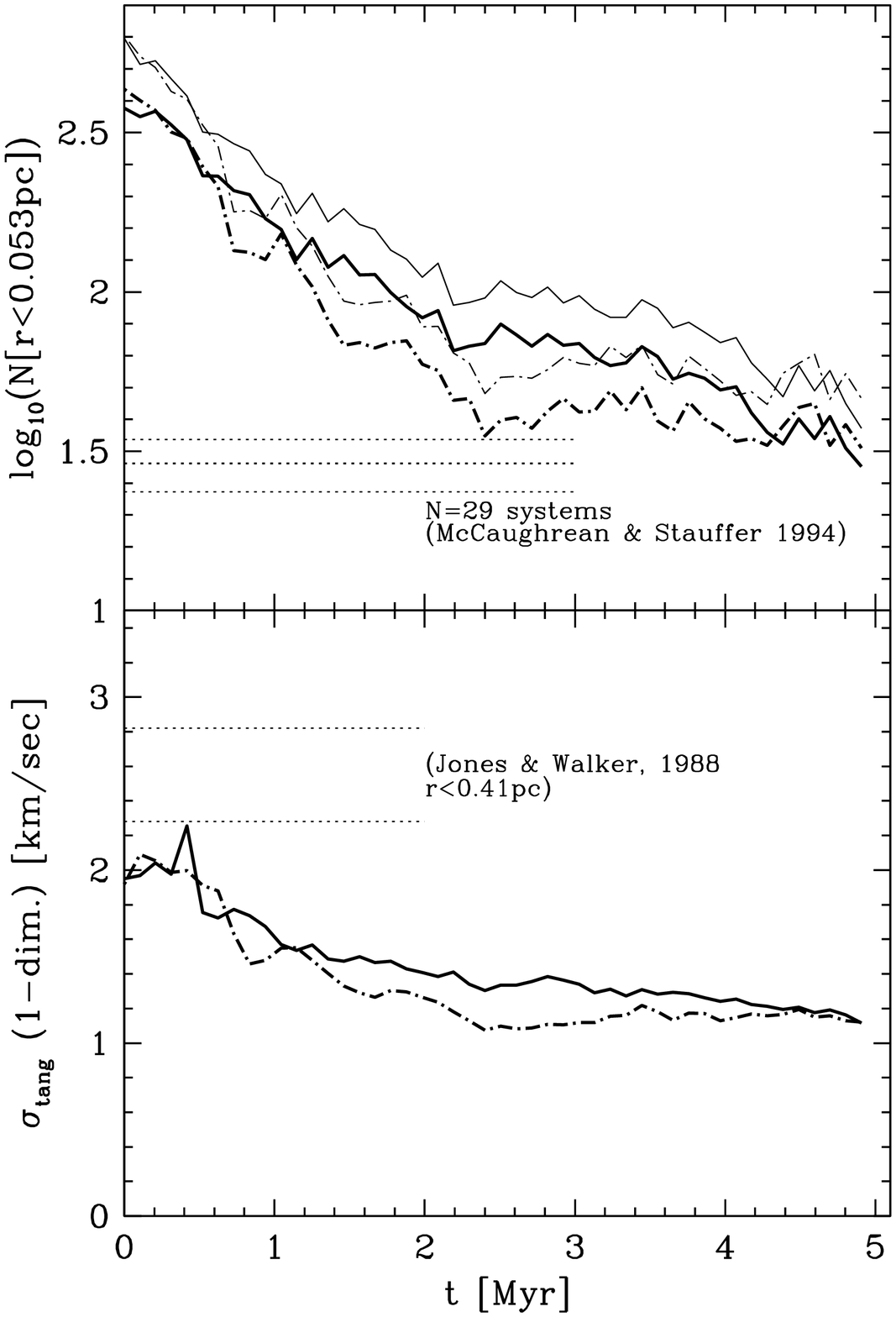,width=3.5in}}
\caption{Upper panel: The time evolution of the number of stellar
systems within 0.053\,pc of the density maximum of the model cluster.
The thick curves assume no binary systems are resolved, and the thin
curves count all stars.  The observational constraint with the Poisson
error range is indicated by the dotted lines. Lower panel: The time
evolution of the one-dimensional velocity dispersion of
centres-of-mass within a projected distance of 0.41\,pc from the
density maximum of the model cluster. The observational one-sigma
uncertainty range is indicated by the dotted lines.  In both panels,
the solid curves are for Model~A and the dot-dashed curves for
Model~B\@.}
\label{fig:d_vd}
\end{figure}


\subsection{A model Trapezium Cluster: overall evolution} \label{cl_evol}
From ground-based near-infrared images of the Trapezium Cluster core, 
McCaughrean \& Stauffer (1994) estimate that 29~systems are within the 
central spherical volume with a radius of 0.053\,pc. 

Figure~\ref{fig:d_vd} compares the evolution of the number of systems
within the central region of the model clusters with the above
observational constraint. The models are initially over-dense by an
order of magnitude, but agree with the observational constraint after
2.5\,Myr (Model~B) and 4.2\,Myr (Model~A). The cluster expands mostly
through three- and four-body interactions. In a cluster consisting
initially of single stars, on the other hand, the central number
density increases continuously (\eg{} Spurzem \& Takahashi 1995), until
the core collapses at $\approx$\,2--3\,$\tr$ (Spitzer 1987). That the
number of systems is consistently higher in Model~A than in Model~B is
is due to enhanced binary destruction in Model~A, which liberates
new single stars. The disruption of wide binaries also cools the
cluster which slows the expansion for Model~A relative to Model~B,
leading to a larger density in the core.

%
\begin{figure}[t]
\centerline{\psfig{figure=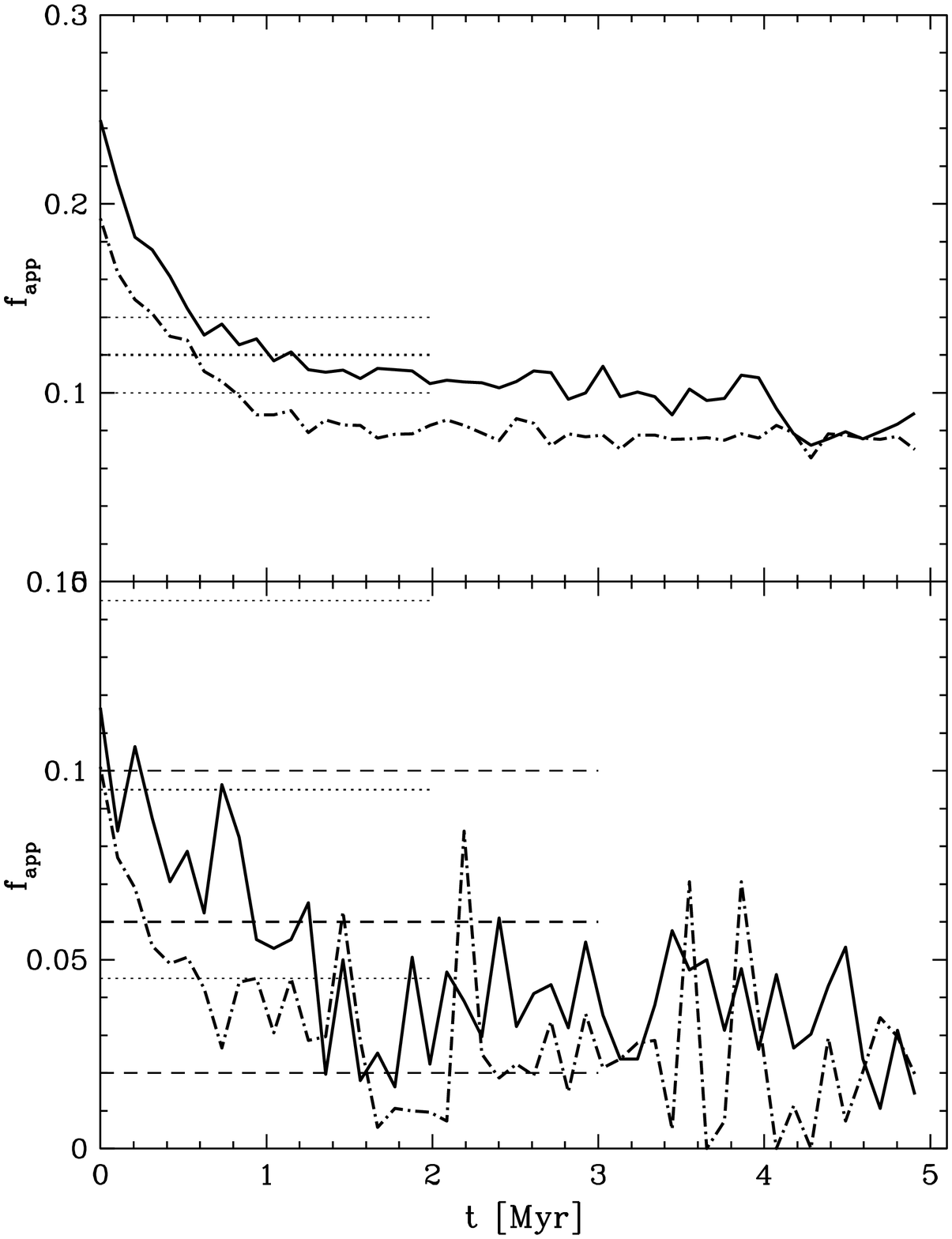,width=3.5in}}
\caption{The time evolution of the apparent binary proportion in the
model Trapezium Clusters. Models~A and~B are the solid and dot-dashed
lines respectively. Upper panel: observational constraints from
Prosser \etal{} (1994) are shown as dotted lines. Lower panel:
observational constraints from Petr \etal{} (1998) are shown as the
horizontal lines. Here, the central dotted line is $f_{\rm app}$ for
all systems in their sample, and the central dashed line is $f_{\rm app}$ 
for the low-mass systems only. Poisson uncertainties are
indicated by the upper and lower horizontal lines.}
\label{fig:fapp}
\end{figure}


The one-dimensional tangential velocity dispersion within 0.41\,pc of
the centre of the Trapezium Cluster, is estimated by Jones \& Walker
(1988) to be $2.54\pm0.27$~km/s from their proper-motion survey.  The
data indicate no anisotropy, but these authors, van Altena \etal{} (1988),
and Tian \etal{} (1996) note that the plate-reduction algorithms used 
eliminate any signature due to rotation, expansion, or contraction. 
Thus, it is presently unknown if the Trapezium Cluster is expanding 
or contracting.

The evolution of the one-dimensional velocity dispersion within
0.41\,pc of the density maximum of the model clusters is shown in
Figure~\ref{fig:d_vd}. The velocity dispersion increases slightly during
the first few $10^5$\,yr because the numerical model contracts slightly
during adjustment to viral equilibrium, which is never perfectly
achieved in a discrete rendition of a dynamical system. The velocity
dispersion is similar to the observed value during these times, but
then decreases substantially. The velocity dispersion is inconsistent
with the observational constraint when the central number density
agrees with the observations.

\subsection{A model Trapezium Cluster: binary stars} \label{bins-evol}
The observations in the core of the Trapezium Cluster by Petr \etal{}
(1998; this volume) resolve binary systems with separations
over the range 63--225\,AU\@. Of the 42 systems that appear projected
within the central radius of 0.041\,pc, six are OB stars and four are
binary systems. The apparent binary proportion of the entire sample
is $f_{\rm app}=0.095\pm0.05$, and the binary proportion for low-mass
($m<1.5$\Msolar) stars is $f_{\rm app,lms}=0.06\pm0.04$.  The
subscript ``app'' means that the observed $f$ is the apparent binary
proportion that an observer deduces from projected star positions
within some range of separations to which the observational apparatus
is sensitive. The binary proportion of low-mass stars within a radius
of 0.25\,pc of the centre of the Trapezium cluster was found by Prosser
\etal{} (1994) to be $f_{\rm app}=0.12\pm0.02$ for separations over the
range 26--440\,AU\@.

Observations and theory are compared in Figure~\ref{fig:fapp}.
Significant evolution of the model binary population occurs within the
first 1\,Myr, and $f_{\rm app}^{\rm Model~A}>f_{\rm app}^{\rm Model~B}$
in the upper panel, suggesting that the initial binary proportion and
period distribution can, in principle, be constrained. The above
inequality is violated for times $t>1$\,Myr in the much smaller sample
in the cluster core, and information on the initial binary proportion
is lost. The model results show that the observational constraints
are still too weak to allow a distinction between Models~A and~B. Both
are consistent with the data for dynamical cluster ages $t>0.3$\,Myr.

\begin{figure}[t]
\centerline{\psfig{figure=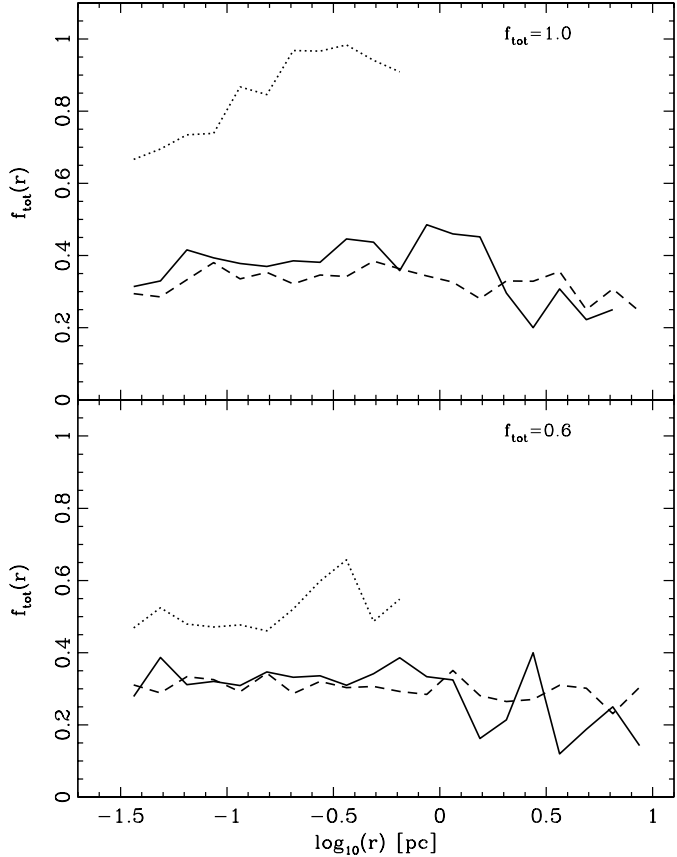,width=3.5in}}
\caption{Projected radial dependence of $f_{\rm tot}$.  In both the
upper panel (Model~A) and the lower (Model~B) panel, the dotted line 
is the initial distribution, and the solid and dashed lines are at 1\,Myr
and 5\,Myr, respectively.}
\label{fig:ftot_rad}
\end{figure}

The observations by Petr \etal{} (1998; this volume) and Prosser \etal{}
(1994) suggest that $f$ is roughly independent of radial distance.
As is evident from Figure~\ref{fig:ftot_rad},
initially there are fewer binaries near the centre. This is a result
of disruption from crowding, and is much more pronounced for Model~A,
which is more abundant in long-period binaries than Model~B\@. After
1\,Myr, there remains a slight radial dependence, with $f_{\rm tot}$
increasing slightly with increasing $r<1.3$\,pc in Model~A. The decay
for larger $r$ is a result of primarily low-mass stars being expelled
to large radii after three- and four-body interactions near the
cluster core.  The radial dependence has vanished at $t=5$\,Myr, and
$f_{\rm tot}\approx 0.35$.  In Model~B, $f_{\rm tot}\approx 0.3$ for
$t>1$\,Myr and all $r$, and little further evolution is apparent until
$t=5$\,Myr.

%
%

The models confirm that $f_{\rm tot}$ should be independent of $r$ for
$t>1$\,Myr, by which time the stellar population is well mixed. If
observations find $f(r\approx1.3\,{\rm pc})>f(r>1.3\,{\rm pc})$ and/or
$f(r\approx1.3\,{\rm pc})>f(r<0.3\,{\rm pc})$, then this would be
evidence for $f_{\rm tot}\approx 1$ at birth, and that the Trapezium
Cluster is merely on the order of one crossing time old.

\subsection{A model Trapezium Cluster: eccentricity-period diagram} 
\label{ecc-P}
\begin{figure}[t]
\centerline{\psfig{figure=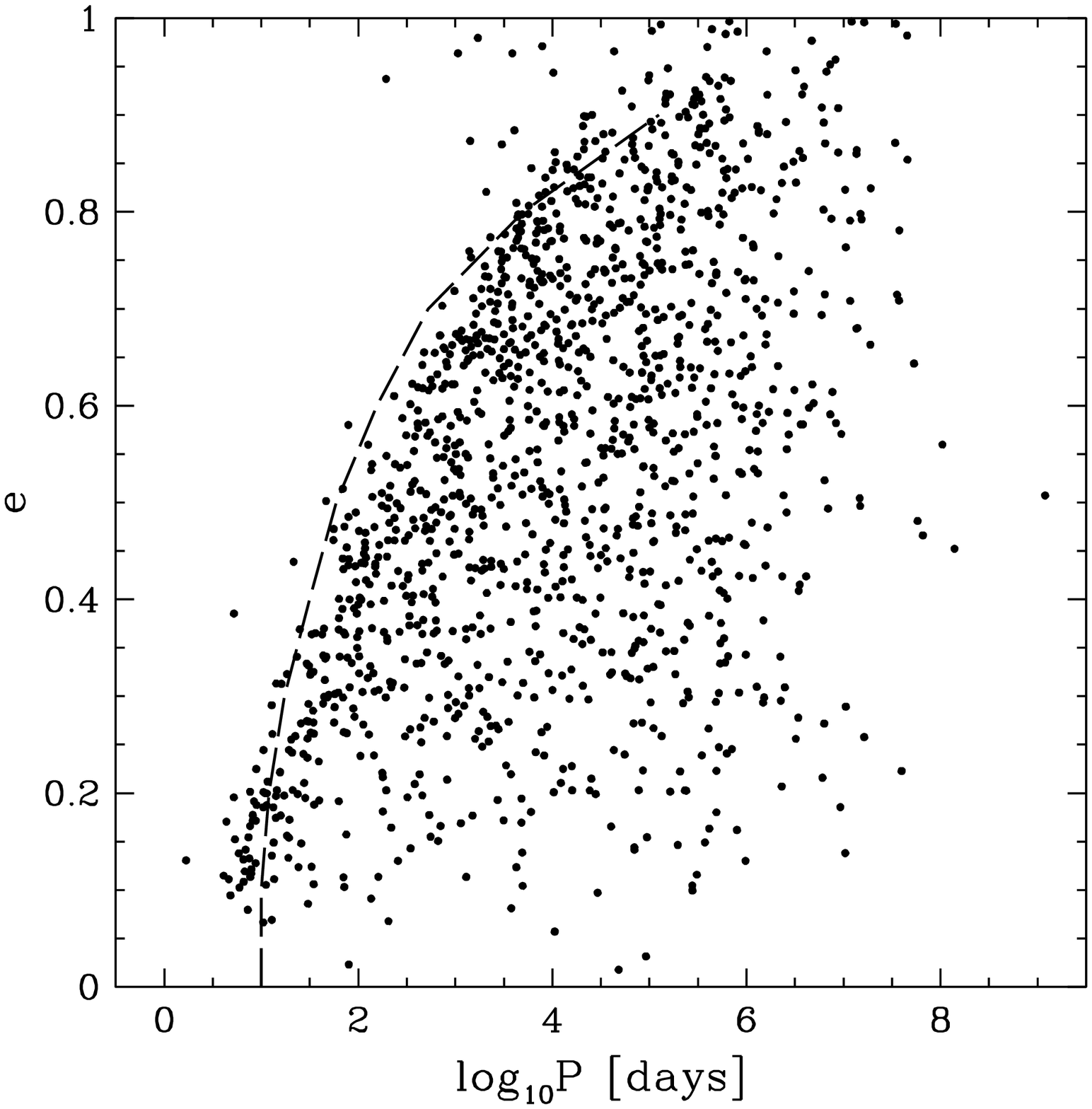,width=4in}}
\caption{The eccentricity-period diagram for Model~A at time
$t=1$\,Myr. All surviving orbits of the three simulations are plotted.
The distribution for Model~B is very similar.  The thick dashed line
represents the observed envelope for main-sequence binary stars with a
G-dwarf primary (Duquennoy \& Mayor 1991). The initial model data are
distributed to the right of the envelope (see Figure~5 of Kroupa 1995b).
Systems with eccentricities above this line with $e$ of order 0.2
or greater are therefore evidence for stellar encounters.}
\label{fig:e_p_1Myr}
\end{figure}

The dynamical interactions between stellar systems near the core of a
cluster have a rich variety of outcomes.  Binary systems that survive
such an encounter will have their orbital properties significantly
affected. The relative change in
eccentricity is usually much larger than the relative change in
binding energy (Heggie 1975). A binary system emanating from an
interaction will often have an eccentric orbit, and may be
identified in the eccentricity-period diagram if it lies outside the
observed envelope (see Figure~\ref{fig:e_p_1Myr}).

%
%

The eccentricity-period diagram after 1\,Myr is shown for Model~A in
Figure~\ref{fig:e_p_1Myr}.
The region to the left of the envelope has been populated. Those
binaries with a very large eccentricity would lead to physical
collisions between the companion pre-main sequence stars. However, it
is noteworthy that the two orbits furthest away from the envelope, (a)
$e=0.39, \log P=0.72$ and (b) $e=0.94, \log P=2.29$,
are systems that have been ejected from the cluster.  System~(a)
consists of two stars each with a mass of 2.79\Msolar, at a
distance of 8.4\,pc from the model cluster and receding with 50.3\kmpers{}
from it. System~(b) is located at a distance of 13.1\,pc from the
cluster, is receding from it with a velocity of 22.8\kmpers, and consists
of a 0.51\Msolar{} and a 0.35\Msolar{} star. Real systems with
such large eccentricities and short orbital periods will undergo
significant orbital evolution within probably 10--1000 orbital
periods, \idest{} within less than a few thousand years. How fast tidal
circularisation proceeds remains uncertain (see Mathieu 1994; Kroupa
1995b and references therein; Verbunt \& Phinney 1995), given that the
distended internal structure of the pre-main sequence stars is an
important ingredient of tidal circularisation theory, but it is likely
to slow down as circularisation progresses. An observer is therefore
most likely to capture such systems towards the end of the
circularisation phase, when the eccentricity has decayed considerably.
Possible observational counterparts to such systems are found
in Table~A2 in Mathieu (1994): P2486 has spectral class G5, 
has $e=0.161$ and $\log P=0.72$, and is located in the Trapezium Cluster; 
OriNTT~429 is a K3 system with $e=0.27$ and $\log P=0.87$, but does 
not appear to be in a young cluster.


\section{Circumstellar Discs}

\begin{figure}[t]
\centerline{\psfig{figure=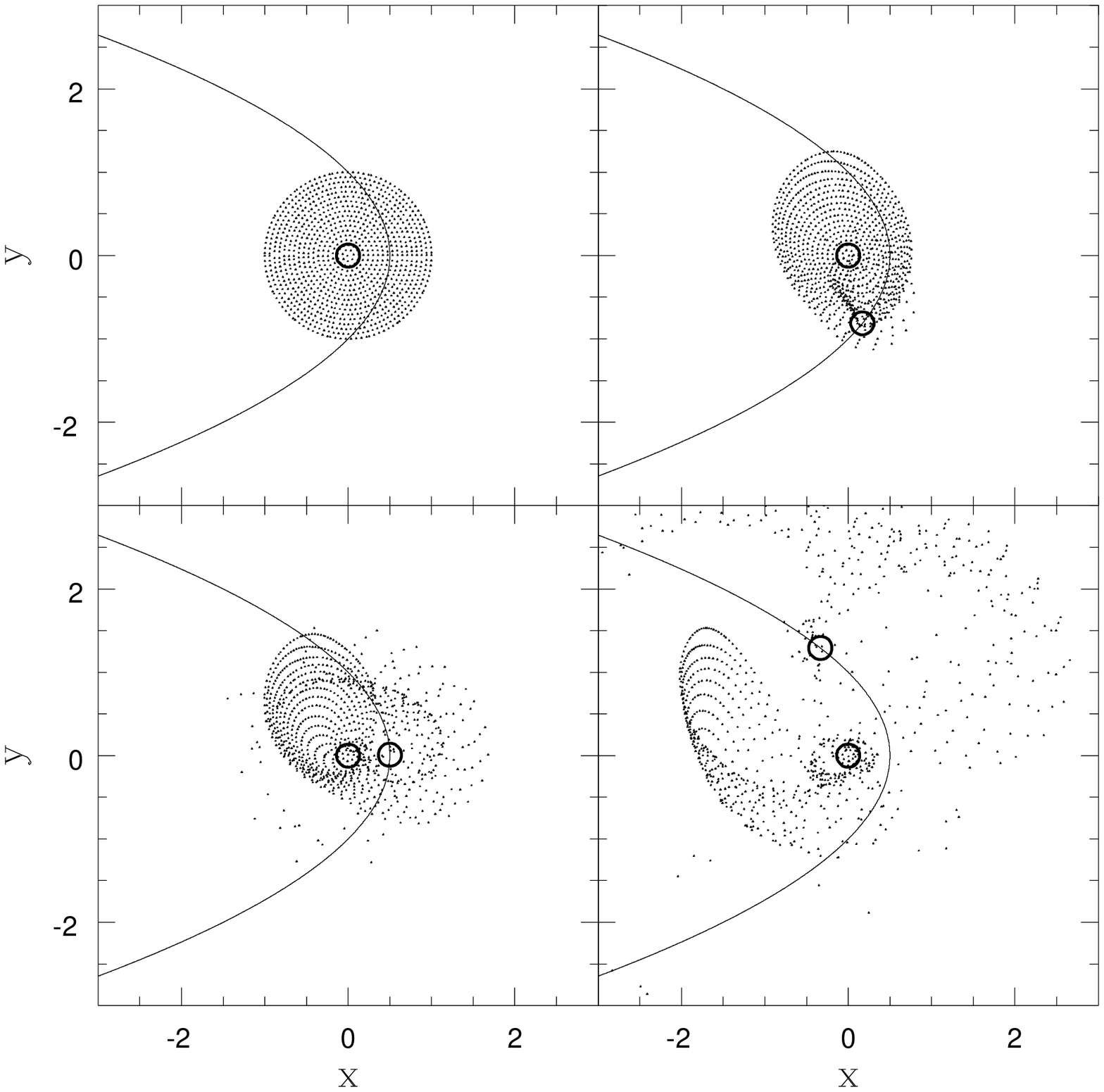,width=4in}}
\caption{\label{star_disk} A star-disk interaction where the disk
matter exterior to periastron is unbound in tidal tails (Hall, Clarke,
\& Pringle 1996).}
\end{figure}

Stellar encounters involving single stars and their circumstellar
discs are also an important dynamical process in stellar clusters as
their large sizes (100--1000\,AU) result in significant cross
sections for interactions.  Circumstellar discs are a ubiquitous
by-product of the star formation process due to the presence of even
small amounts of angular momentum in the pre-collapse molecular
clouds. These discs are commonly found around young stars of ages
$\simless$\,few Myr (Strom, Edwards, \& Strutskie~1993). Interactions 
between stars and discs in a clustered environment can play an important 
role in the dynamics of the stars and of the discs.

Star-disc interactions have been proposed as a mechanism to form
binary systems (Larson 1990; Clarke \& Pringle 1991; see Clarke 1995),
by way of removing kinetic energy from the stellar motions and leaving
the system bound.  The necessary encounter is a violent one, with the
perturber's orbit passing through the other star's disc, unbinding the
matter initially outside of the periastron passage.  Star-disc
interactions can be effective in small clusters where the velocity
dispersion is low (Clarke \& Pringle 1991; McDonald \& Clarke 1995;
Clarke 1995; Hall, Clarke, \& Pringle 1996), and can help explain the
high incidence of binary systems amongst pre-main sequence stars such
as is found in Taurus (Ghez, Neugebauer, \& Matthews 1993; Leinert \etal{} 
1993). This mechanism may also lead to a high binary proportion in dense 
clusters if fragmentation initially produces sub-clumps, each of which is a
small cluster.

In larger clusters and after the possible initial sub-clumping is
erased, the velocity dispersion is higher and the chance of a
destructive high velocity encounter before a low velocity, and hence
capturing, encounter is large (Clarke 1995).  Thus, in clusters such
as the ONC, the primary effect of star-disc interactions is the
disruption they inflict on the discs and thus on the disc's evolution.
The encounter will generally remove all disc material exterior to the
periastron separation (Figure~10). This paring down of the disc results
in a smaller disc mass and a disc radius of $\approx 1/3$ of the
original periastron radius (Clarke \& Pringle 1993; Heller 1995; Hall
\etal{} 1996; Larwood 1997). Additionally, the disc displays an
exponential outer edge similar that observed for several of the
silhouette disks in the Trapezium Cluster (Hall 1997; 
McCaughrean \& O'Dell 1996; see also the chapter by McCaughrean
\etal{} in this volume).

If the original periastron radius is not too small, the disc
compression caused by the encounter can decrease the formation
timescale of outermost planets. This may lessen the discrepancy
between the excessively long formation timescale of Neptune and
empirical proto-stellar disc life-times if the Sun formed in an
embedded cluster (Eggers \etal{} 1997), which is also corroborated by
the obliquity of the planetary system (Heller 1993).


\section{Gas Expulsion}
Young clusters contain significant amounts of gas, typically
comprising a majority of the total cluster mass (Lada 1991). This gas
is therefore a major contributer to the cluster potential and its
removal can unbind the cluster, dispersing the stars into the Galaxy.
Gas removal in clusters occurs due to the energetics of the component
massive stars (Whitworth 1979; Tenorio-Tagle \etal{} 1986; Franco, Shore,
\& Tenorio-Tagle 1994). The photo-ionization and winds from these
stars is capable of removing any residual gas from the cluster.

The fate of a particular cluster depends on the gas fraction, the
removal timescale and stellar velocity dispersion when the gas is
dispersed (Lada, Margulis, \& Dearborn{} 1984; Pinto 1987; Verschueren
\& David 1989; Goodwin 1997b; Saiyadpour \etal{} 1997). If the
gas is removed quickly compared to the cluster crossing time, \tcr,
then the dramatic reduction in the binding energy, without affecting
the stellar kinetic energy, results in an unbound cluster for any
reasonable gas mass fraction, unless the stellar cluster was in a
state of collapse prior to gas removal.  Alternatively, if the gas is
removed over several crossing times, then the cluster can adapt to the
new potential and can survive with a significant fraction of its
initial stars. For example, clusters with gas fractions as high as 80\%
can survive with approximately half of the stars if the gas
removal occurs over four or more crossing times (Lada \etal{} 1984).
Clusters with larger central concentrations survive preferentially as
the cluster core with a high stellar density is less affected by the
removal of gas from the rest of the cluster (Lada \etal{} 1984;
Goodwin 1997b). However, violent gas expulsion through massive
outflows from the massive central stars (Churchwell 1997) is likely to
be an important process governing core evolution.

The number and age distribution of Galactic clusters suggests that
only a few percent of all Galactic field stars can have originated in
bound clusters (Wielen 1971). However, star counts in molecular clouds
(Lada \& Lada 1991) and the properties of Galactic field binaries
(Kroupa 1995a) indicate that most stars may form in clusters. The
implication is that the life-time of the typical embedded cluster is
$\simless$\,10\,Myr (Battinelli \& Capuzzo-Dolcetta 1991), which is
a natural consequence of rapid gas expulsion and low local star
formation efficiency.

\section{Summary}
The formation of a stellar cluster involves many dynamical processes
that need to be understood. The least understood of these is the
initial formation mechanism from which hundreds to thousands of stars
form nearly simultaneously in a bound grouping. Subsequent evolution
depends to a large extent on interactions with, and accretion of, the
gas by the stars. Competitive accretion in a cluster potential results
in a large range of stellar masses, comparable to that observed in the
Orion Nebula Cluster. It is thus a strong candidate process to explain the
observed distribution of stellar masses. This process naturally
results in forming the most massive objects near the cluster centre
whereas their location there cannot be due to dynamical mass
segregation. Accretion also shrinks the stellar distribution, such
that with sufficient accretion, it is possible for collisions to occur
in the cluster core, thereby aiding in the formation of massive stars.

Stellar interactions in young clusters are significant, as they
commonly have circumstellar discs, and are often in binary and
multiple systems, thereby increasing their cross section for such
interactions. Binary-single and binary-binary interactions in young
clusters are generally destructive, unbinding the wider systems and
reducing the binary fraction. Assuming that the normal mode of star
formation is that of binary stars (as is found in Taurus-Auriga), then the
observed binary fraction in different stellar populations is an
indication of the subsequent dynamical interactions and thus the
degree to which the star formed in a clustered environment. The
observed binary fraction in the Trapezium Cluster is consistent with
having originally been as high as that found in Taurus-Auriga, and having
subsequently been reduced through stellar interactions. If this is the
case, then the binary fraction in the outer, less evolved part of the
cluster, should be as high as that in Taurus-Auriga.

Interactions between multiple- and single-star systems 
in an initially binary-rich cluster significantly
affect the early evolution of the cluster. The core expands with a
rate depending on the number ratio of hard to soft binaries. In
single-star clusters, on the other hand, the core will contract.

Star-disc interactions in small clusters and sub-clusters can aid in
the formation of binary systems by transferring kinetic energy from
the stars' orbits to unbinding parts of the disc.  In larger clusters,
these interactions are more violent and will rarely form binary
systems but will significantly affect the discs, removing that
fraction of their mass located beyond periastron. Disc compression
during less-destructive star-disc encounters can induce planet
formation and decrease the timescale for assembly of outer planets.

The majority of young clusters will not form bound open clusters but
will instead disperse their stars into the field. This must be the
case if most stars are formed in clusters but spend the majority of
their lifetime as field stars. The cluster dispersal will generally
occur when either---for small clusters---an evaporation time has been
reached or---for larger clusters---when a massive star has removed all
remaining gas and thus unbinds the cluster.


\acknowledgments We thank the editors and the organisers of the
Ringberg conference on the Orion star formation complex for having
invited us to a stimulating meeting.  Sverre Aarseth is thanked warmly
for readily providing his {\sc Nbody5} code.


\end{document}